\documentclass[aps,prl,preprint,superscriptaddress,amsmath,amssymb]{revtex4}
\usepackage{graphicx}
\usepackage{color}
\begin{document}
\title{Effect of intense, ultrashort laser pulses on DNA plasmids in their native state: strand breakages induced by {\it in-situ} electrons}
\author{J. S. D$'$Souza}
\affiliation{Centre for Excellence in Basic Sciences, University of Mumbai, Kalina Campus, Mumbai 400 098, India}
\author{J. A. Dharmadhikari}
\affiliation{Tata Institute of Fundamental Research, 1 Homi Bhabha Road, Mumbai 400 005, India}
\author{A. K. Dharmadhikari} 
\affiliation{Tata Institute of Fundamental Research, 1 Homi Bhabha Road, Mumbai 400 005, India}
\author{B. J. Rao}
\affiliation{Tata Institute of Fundamental Research, 1 Homi Bhabha Road, Mumbai 400 005, India}
\author{D. Mathur}
\email{atmol1@tifr.res.in}
\affiliation{Tata Institute of Fundamental Research, 1 Homi Bhabha Road, Mumbai 400 005, India}

\begin{abstract}
Single strand breaks are induced in DNA plasmids, pBR322 and pUC19, in aqueous media by intense ultrashort laser pulses (820 nm wavelength, 45 fs pulse duration, 1 kHz repetition rate) at intensities of 1-12 TW cm$^{-2}$. The intense laser radiation generates, {\it in situ}, electrons that induce transformation of supercoiled DNA into relaxed DNA. The extent of electron-mediated relaxation of DNA structure is quantified. Introduction of electron and radical scavengers inhibits DNA damage. 
\end{abstract}
\maketitle

Studies of the interaction of radiation with biological matter have long focused on quantification of the energy that gets transferred from the radiation field into irradiated matter. The physics governing primary energy deposition has been understood for some decades \cite{inokuti} and finds applications in the biomedical sciences. Information now exists that readily enables not only deduction of macroscopic patient doses in radiotherapy \cite{Horton} but also microdosimetric doses within single cells \cite{rossi}. However, while it is now routinely possible to quantify the energy that is deposited in a given volume of irradiated matter, there remains a gap in knowledge as to the subsequent sequence of events that unfold. As a result quantitative insights into how a given dose of radiation induces biological effects continue to be elusive. Since the most important radiation damage is that caused to the genome, it is clear that the focus of experimental studies must be on DNA. The breaking of single and double DNA strands constitutes potentially the most lethal damage at the cellular level. For long it has been thought that such damage is caused by exposure of living matter to high-energy radiation \cite{radiation_damage} that ionizes the sugar-phosphate backbone. However, about a decade ago, Sanche and coworkers pioneered a new line of gas-phase experiments that offered indications that even electrons possessing only a few eV's of energy might induce single strand breaks (SSB) and double strand breaks (DSB) \cite{sanche_early} through the formation of temporary negative ion states that subsequently dissociate.  

Breakage of DNA strands by low energy electrons is of interest as such electrons are copiously produced along tracks of ionizing radiation, typically about 10$^4$ electrons per MeV that is deposited \cite{radiation}. Li {\it et al.} \cite{Li} carried out model calculations in which a section of DNA backbone was represented by two deoxyribose (sugar) rings that were connected by a phosphate; {\it ab initio} computations of adiabatic potential energy surfaces of the neutral and the anion revealed that bond rupture is thermodynamically favorable as activation barriers are exceedingly low ($\sim$10 kcal mol$^{-1}$). Hence, interactions involving even thermal energy electrons would be expected to lead to bond cleavages. In solution phase, the energetics are likely to be somewhat different because of solvation effects. Even though the reaction between a solvated electron and dialkyl phosphate anions proceeds slowly, results of computations have indicated that direct damage to the DNA backbone by low energy electrons may be so fast that it actually precedes solvation \cite{Li}. Indeed, an electron with only about 0.5 eV energy would lead to formation of DNA multiple transient anion states which decay into damaged structures involving SSB and DSB \cite{Huels}.

We report here results of experiments that we have conducted to explore electron-mediated damage to DNA in its native, aqueous state. The electron-mediated damage manisfests itself in the creation of relaxed forms of DNA which we monitor using gel electrophoresis. We observe that upon addition of electron scavengers like 5-bromo-uracil, there is a significant reduction in the population of relaxed species. Similar reduction is obtained upon addition of melatonin, a scavenger of free radicals. Damage is, therefore, essentially caused by electrons and free radicals that are produced in the course of the intense laser's interaction with aqueous water+DNA. The electrons are produced {\it in situ} by ultrashort (45 fs duration) pulses of 820 nm light of peak intensities ($I$) in the range 1-12 TW cm$^{-2}$ at a repetition rate of 1 kHz. Descriptions of the appratus can be found in our recent reports on supercontinuum generation in materials like BaF$_2$ and in macromolecular media \cite{whitelight}. The intensity of light that is incident on the DNA+water sample is high enough for us to invoke the optical Kerr effect wherein the total refractive index ($n$) comprises a linear and an intensity-dependent nonlinear portion, $n = n_o+n_2I$. The laser beam's Gaussian intensity profile then maps to a refractive index profile $n = n_o+n_2I~exp(-2r^2/w_o^2)\approx n_o+n_2I(1- 2r^2/w_o^2)$. The radial dependence of the phase of the propagating beam results in self-focusing within the irradiated aqueous medium until high enough intensity is attained for multiphoton ionization (MPI) to occur. MPI-generated electrons, in turn, contribute to de-focusing such that propagation through the medium proceeds in a series of self-focusing-de-focusing events  that have been recently visualized by means of 6-photon absorption-induced fluorescence in BaF$_2$ \cite{filamentation} (for a cogent review, see \cite{physrep}). The plasmid DNA (pUC19 and pBR322) used in our work are from a commercial source (Bangalore Genei, India). These plasmids were suspended in 2 liters of de-ionized water in dialysis bags with a molecular size cut-off of 12 kDa. Changes were made twice every 3 hours after which they were dispensed into convenient volumes and stored in Eppendorf tubes at -20 C. The concentration of the DNA was  spectrophotometrically measured at 260/280 nm wavelengths. Supercontinuum spectra of pure water and DNA+water were measured with a fiber-based spectrometer over the spectral range 200-870 nm; asymmteric broadening towards the blue of the supercontinuum provided ready confirmation of plasma formation within the irradiated medium \cite{whitelight}. Post irradiation, electrophoresis enabled separation of DNA fragments by size. After separation, the gel was stained with a DNA binding fluorescent dye, ethidium bromide, to enable viewing by a BIORAD Gel Documentation system. Fragment size determination was accomplished with reference to commercially available DNA ladders containing linear fragments of known length. Since electrophoresis techniques are used for assessment of DNA damage, the documentated gels were further used for measurement and analysis using standard gel-analysis software (ImageJ). 

Typical data for percentage change in supercoiled DNA upon irradiation with our laser pulses is shown in Fig. 1. Is the dramatic increase that is observed in fraction of relaxed DNA induced by energetic photons? Figure 1 also depicts typical supercontinuum spectra that we measured. The bluest part of the supercontinuum is seen to clamp at $\sim$400 nm; the absence of 266 nm radiation rules out single-photon damage to DNA. Furthermore, 2- and 3-photon absorption from the supercontinuum is negligible. Gel electrophoresis allows us to quantify the extent of DNA damage on important parameters like the time for which the sample is irradiated (akin to the radiation dose), the DNA concentration, and the laser energy. 

\begin{figure}
\includegraphics[width=10cm]{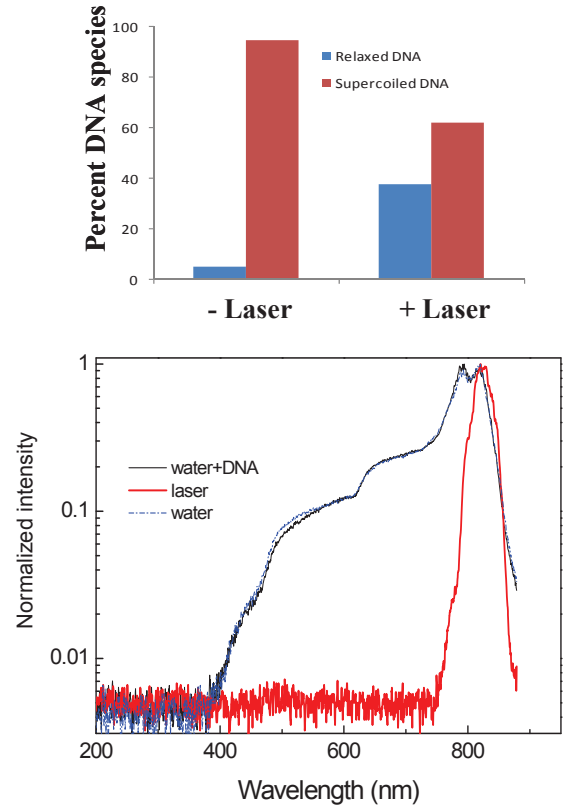}
\caption{(Color online) The top panel shows the percentage of supercoiled and relaxed DNA in normal conditions (- Laser) and after irradiation (+ Laser). The lower panel depicts spectra of white light generated upon irradiation of water and water+DNA with 820 nm light of intensity 5 TW cm$^{-2}$. The narrower spectrum of the incident laser beam is also shown. Note the logarithmic scale.}
\end{figure}

Figure 2 shows data measured at a fixed concentration of DNA, and with the laser energy fixed at 130 $\mu$J, as the exposure time was varied over the range 10-120 seconds. The gel electrophoresis data and the corresponding graphical representation show that the fraction of relaxed species increases to $\sim$15\% at 120 s exposure time. We conducted subsequent experiments using this exposure and data show that as the DNA concentration was varied over the range 2-6 $\mu$g/$\mu\ell$, $\sim$66\% to 15\% of the supercoiled DNA was converted to the relaxed form; there was some dependence on which plasmid our measurements were conducted on. In the case of pBR322 plasmid, 65\% conversion into relaxed species was obtained at a concentration of 3 $\mu$g/$\mu\ell$ while for pUC19, only 45\% conversion was achieved at this concentration. Similarly, intensity dependent data showed a yield of 10-33\% relaxed species as the laser intensity varied over the range 1-4 TW cm$^{-2}$, independent of which plasmid was used. 

\begin{figure}
\includegraphics[width=10cm]{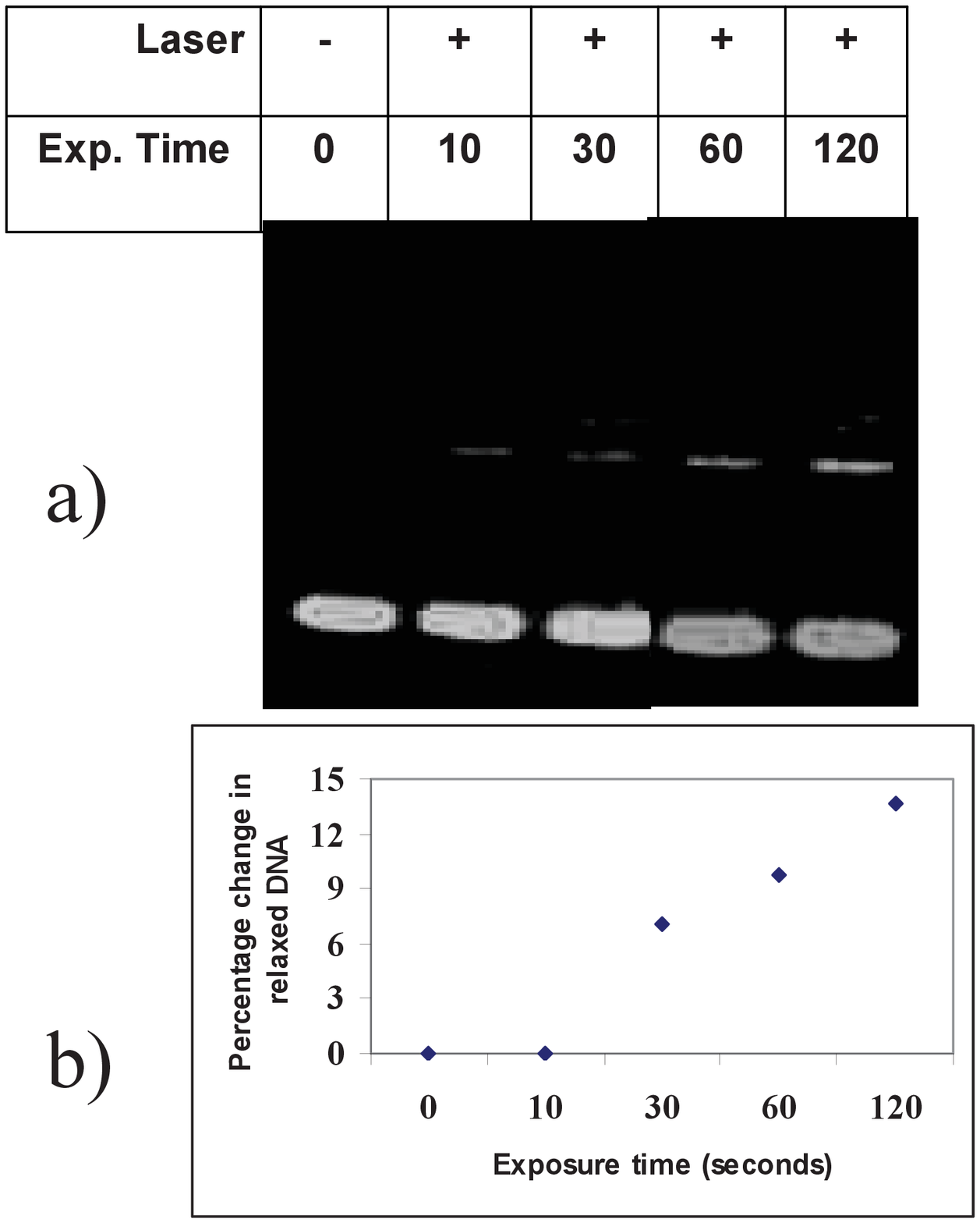}
\caption{a) Gel electrophoresis data for DNA plasmid pBR322 exposed for periods of times ranging from 0 to 120 seconds. The lower bright image denotes supercoild DNA while the upper traces that become weakly visible at 10 s and more prominently visible after 60 s denote relaxed DNA. b) Graphical quantification of the gel data. The DNA concentration was 2 $\mu$g/$\mu\ell$ and the laser intensity was 3 TW cm$^{-2}$.}
\end{figure}

We identify three possible processes that might set in as the intense laser pulse propagates through the DNA+water solution:
\begin{enumerate}
	\item The laser light (800 nm) can be absorbed by the DNA through three-photon absorption (direct process).
	\item Multiphoton ionization of water can generate free electrons which, in turn, react with DNA so as to cause strand breaks via dissociative electron attachment and the formation of transient negative ions.
	\item These free electrons can react with water molecules to form free radicals, like *OH, which, in turn, may also mediate in strand breaking processes (indirect process).
\end{enumerate}

Plasma formation in water that has been irradiated by intense laser light has been well studied (for a review, see \cite{vogel}) and the breakdown process has been modeled \cite{saachi} by treating water to be an amorphous semiconductor with a band gap of 6.5 eV. Moreover, nonlinear absorption of liquid water not only involves ionization but also dissociation of the water molecules, leading to formation of reactive species like *OH. The quasi-free electrons that are produced gain further energy from the optical field via inverse bremsstrahlung and participate in further ionizing collisions. Rate equations for optical breakdown in water indicate that electron densities of 10$^{18}$-10$^{20}$ cm$^{-3}$ can be attained \cite{vogel,shen} and, we postulate, that it is these electrons that contribute to formation of temporary negative ions in water+DNA. The breakup of such negative ions results in strand breakages \cite{Li,Huels}. 

\begin{figure}
\includegraphics[width=10cm]{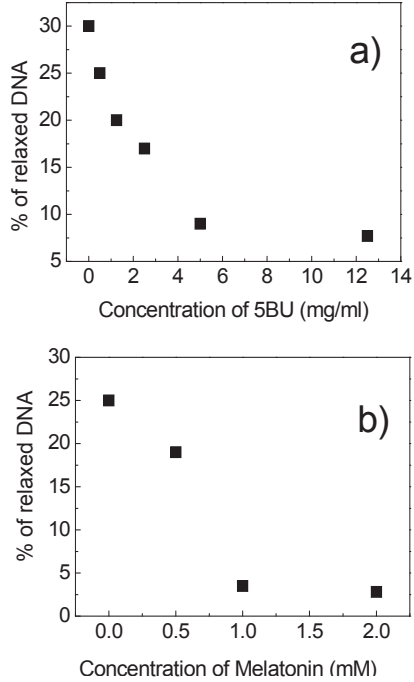}
\caption{Percentage of relaxed DNA as the concentration is increased of a) the electron scavenger and b) the radical scavenger.}
\end{figure}

To ascertain whether the relaxed form of DNA post-laser exposure in our experiments was, indeed, mediated by {\it in-situ} production of free electrons or radicals like OH$^-$ or OH, we carried out experiments wherein effective electron quenchers such as 5-bromouracil or 5-bromo-2,4(1H,3H)-pyrimidinedione) were added to the water-DNA sample. We also made measurements when water+DNA was doped with melatonin (N-acetyl-5-methoxytryptamine), which is a direct scavenger of radicals like OH, O., and NO. These electron and radical scavengers were chosen with care to ensure that they would not chemically react with DNA in such manner as to induce strand breakages. The results we obtained upon such doping are depicted in Fig. 3 and they present clearcut evidence that upon removal of free electrons and/or radicals, the extent of DNA damage is significantly curtailed. 

In summary, our experiments on laser-DNA interactions in the liquid phase have (i) demonstrated a method of generating, {\it in-situ}, electrons and free radicals in an aqueous environment; (ii) such electrons and free radicals interact with DNA plasmids kept in physiologically-relevant conditions so as to produce nicks in the plasmid DNA; and (iii) the number of nicks thus produced is measured to be directly proportional to the time of laser exposure, laser intensity, concentration of the plasmid, and the type of plasmid. From a biological perspective, we note that enzymes such as gyrases and topoisomerases are known to induce nicks in DNA. Our work seems to suggest that they may well do so by electron- and/or free radical-mediated processes.

We gratefully acknowledge very useful discussions with A. Couairon.


\begin{thebibliography}{00}
\bibitem{inokuti}M. Inokuti, Rev. Mod. Phys. {\bf 43}, 297 (1971).
\bibitem{Horton}J. L. Horton, Handbook of Radiation Therapeutical Physics (Prentice-Hall, New Jersey, 1987).
\bibitem{rossi}H. H. Rossi and M. Zaider, Chap. VII in Microdosimetry and its applications (Springer, New York, 1994).
\bibitem{radiation_damage}D. Becker and M. D. Sevilla, Advances in Radiation Biology (Academic Press, New York, 1993) p. 121; C. von Sonntag, The Chemical Basis for Radiation Biology (Taylor and Francis, London, 1987).
\bibitem{sanche_early}B. Boudaiffa et al., Science {\bf 287}, 1658 (2000), Radiat. Res. {\bf 157}, 227 (2000); X. Pan, P. Cloutier, D. Hunting, and L. Sanche, Phys. Rev. Lett. {\bf 90}, 208102-1 (2003), L. Sanche, Eur. Phys. J. D {\bf 35}, 367 (2005).
\bibitem{radiation}J. A. LaVerne and S. M. Pimblott, Radiat. Res. {\bf 141}, 208 (1995).
\bibitem{Li}X. Li, M. D. Sevilla, and L. Sanche, J. Phys. Chem. B {\bf 108}, 19013 (2004), and references therein.
\bibitem{Huels}M. A. Huels, B. Boudaiffa, P. Cloutier, D. Hunting, and L. Sanche, J. Am. Chem. Soc. {\bf 125}, 4467 (2003).
\bibitem{whitelight}A. K. Dharmadhikari {\it et al.}, Opt. Express {\bf 12}, 695 (2004), Appl. Phys. B {\bf 80}, 61 (2005), {\it ibid.} {\bf 82}, 575 (2006), Phys. Rev. A {\bf 76}, 033811 (2007), C. Santhosh {\it at al.}, J. Biomed. Optics {\bf 12}, 020510 (2007), Appl. Phys. B {\bf 99}, 427 (2010).
\bibitem{filamentation}A. K. Dharmdhikari, F. A. Rajgara, and D. Mathur, Appl. Phys. B {\bf 94}, 259 (2009).
\bibitem{physrep}A Couairon and A Mysyrowicz, Phys. Rep. {\bf 441}, 47 (2007).
\bibitem{vogel}J. Noack and A. Vogel, IEEE J. Quantum Electron. {\bf 35}, 1156 (1999).
\bibitem{saachi}C. A. Saachi, J. Opt. Soc. Am. B {\bf 8}, 337 (1991).
\bibitem{shen}Y. R. Shen, The principes of nonlinear optics (Wiley, New York, 1984).

\end{thebibliography}
\end{document}